# Hydrogen embrittlement controlled by reaction of dislocation with grain boundary in alpha-iron

Liang Wan [a,b,*], Wen Tong Geng [b,c], Akio Ishii [b], Jun-Ping Du [b,d], Qingsong Mei [a], Nobuyuki Ishikawa [e], Hajime Kimizuka [b], Shigenobu Ogata [b,d,**]

[a] Hubei Key Laboratory of Accoutrement Technique in Fluid Machinery and Power Engineering, School of Power and Mechanical Engineering, Wuhan University, Wuhan 430072, China.

[b] Department of Mechanical Science and Bioengineering, Osaka University, Osaka 560-8531, Japan.

[c] School of Materials Science & Engineering, University of Science and Technology Beijing, Beijing 100083, China.

[d] Center for Elements Strategy Initiative for Structural Materials, Kyoto University, Kyoto 606-8501, Japan.

[e] Steel Research Laboratory, JFE Steel Corporation, Kawasaki 210-0855, Japan.

* Corresponding author. Address: School of Power and Mechanical Engineering, Wuhan University, Wuhan 430072, China. Tel: +86 027 68776879; fax: +86 027 68776879.

** Corresponding author. Address: Department of Mechanical Science and Bioengineering, Osaka University, Osaka 560-8531, Japan. Tel: +81 06 6850 6197; fax: +81 06 6850 6197.

*E-mail address:* liangwan@whu.edu.cn (Liang Wan), ogata@me.es.osaka-u.ac.jp (Shigenobu Ogata).

# Abstract

Hydrogen atoms absorbed by metals in hydrogen-containing environments can lead to the premature fracture of the metal components used in load-bearing conditions. Since metals used in practice are mostly polycrystalline, grain boundaries (GBs) can play an important role in hydrogen embrittlement of metals. Here we show that the reaction of GBs with lattice dislocations is a key component in hydrogen embrittlement mechanism for polycrystalline metals. We use atomistic modeling methods to investigate the mechanical response of GBs in alpha-iron with various hydrogen concentrations. Analysis indicates that dislocations impingement and emission on the GB can provoke it to locally transform into an activated state with a more disordered atomistic structure, and introduce a local stress concentration. The activation of the GB segregated with hydrogen atoms can greatly facilitate decohesion of the GB. We propose a hydrogen embrittlement model that can give better explanation of many experimental observations.

# 1. Introduction

For a long time, hydrogen has been known to cause the degradation of the mechanical performance of metals, and this degradation is considered as a key technological challenge in many industrial applications (Gangloff and Somerday, 2012). In essence, hydrogen atoms can be up taken into metals under various environments such as hydrogen gas, moist air, sour gas, water, and acidic solutions. These hydrogen atoms can then be distributed in the lattice via fast diffusion or trapped at defects like vacancies, dislocations, and GBs (Gangloff and Somerday, 2012; Nagumo, 2016). Despite decades of intensive studies, the micro-scale mechanisms of the hydrogen-induced mechanical failure of materials remain a subject of debate (Gerberich, 2012; Kirchheim, 2010; Li et al., 2015; Li et al., 2017; Lynch, 2011; Nagumo, 2016; Neeraj et al., 2012; Novak et al., 2010; Robertson et al., 2012; Robertson et al., 2015; Shibata et al., 2017; Song and Curtin, 2013, 2014; Xie et al., 2016).

The hydrogen-enhanced decohesion (HEDE) theory posits that H atoms gathering at locations of high triaxial stress will lead to the weakening of the bonds of metal atoms and cause fracture (Gerberich, 2012; Oriani, 1987; Troiano, 1960). Although theoretical calculations show that H atoms dissolved in metals can indeed weaken the chemical bonds of the host atoms (Geng et al., 1999; Itsumi and Ellis, 1996) and also a reduction in the cohesive energy of GBs can be induced by the presence of H atoms (Geng et al., 1999; Huang et al., 2017; Kirchheim et al., 2015; Solanki et al., 2012; Wang et al., 2016), it raises important questions: (i) what is the critical level of hydrogen concentration required to induce the fracture; (ii) can this level of hydrogen concentration be achieved under the conditions typically associated with hydrogen-induced failure (Gerberich, 2012; Robertson et al., 2015; Wang et al., 2016); (iii) how can the experimentally observed quasi-cleavage like fractograph of hydrogen embrittled metals (Martin et al., 2011a; Martin et al., 2011b; Nagao et al., 2012) be explained by the HEDE theory.

The observation that dislocation motions can be accelerated under hydrogen charging prompted the suggestion that a stress shielding effect was induced by the hydrogen atmosphere around the dislocations; the local plasticity can therefore be enhanced, which ultimately leads to the early fracture, hence the hydrogen enhanced localized plasticity (HELP) theory (Birnbaum and Sofronis, 1994; Robertson et al., 2012; Robertson et al., 2015). The HELP mechanism is supported by the remarkable experimental facts that well-evolved dislocation structures can be found just beneath the fracture surfaces of hydrogen-embrittled samples, irrespective of whether they are quasi-cleavage-like or intergranular (Martin et al., 2011a; Martin et al., 2011b; Martin et al., 2012; Robertson et al., 2012; Robertson et al., 2015; Wang et al., 2014). However, it has been shown recently by both in-situ experiment (Xie et al., 2016) and atomistic modeling (Song and Curtin, 2014; Zhu et al., 2017) that dislocation motions can sometimes be hampered by hydrogen atoms in the material, which makes the HELP mechanism arguable. Moreover, it remains unclear how local plasticity can be enhanced by an acceleration or hindering of dislocation motion.

There is abundant evidence suggesting that GBs can play an important role in hydrogen embrittlement of polycrystalline metals in the following three aspects. (i) Hydrogen atoms tend to segregate on GBs (Aoki et al., 1994; Mohtadi-Bonab et al., 2013). (ii) For materials charged with fairly high hydrogen concentration, intergranular fracture usually happens (Liu et al., 2008; Martin et al., 2012; Wang et al., 2014); for materials with moderate hydrogen concentration, the quasi-cleavage-like fracture prevails, while the micro-scale facets on the quasi-cleavage fracture surface and also the sites of micro-cracks within the material frequently correspond to GBs (Nagao et al., 2012; Shibata et al., 2015; Shibata et al., 2012). (iii) Suppression of hydrogen embrittlement can be achieved by a refinement of the grain structures (Bai et al., 2016; Macadre et al., 2015; Takasawa et al., 2012). Based on these facts, and in order to clarify the hydrogen embrittlement mechanism, here we have employed atomistic modeling and computational methods to investigate the hydrogen effect on the mechanical response of individual GBs in $\alpha$-Fe. Both a large bicrystal atomistic model using a high fidelity embedded atom method (EAM) interatomic potential for Fe-H and a small bicrystal atomistic model with the first-principles density functional theory (DFT) method were used for the study.

## 2. Computational methodology

2.1 Molecular dynamics (MD) tensile simulation with large bicrystal models

We have first performed MD tensile simulation of GBs with the large bicrystal models. The MD simulations were performed by using the LAMMPS code (Plimpton, 1995). The EAM potential fitted by Ramasubramaniam et al. (Ramasubramaniam et al., 2009), with further modification of the H-H interaction by Song and Curtin (Song and Curtin, 2013), was employed in this study. The Fe-Fe interaction in this potential is the same as the EAM potential for pure iron fitted by Mendelev and co-workers (Ackland et al., 2004; Mendelev et al., 2003), which was shown to be able to give good description of the mechanical properties of α-Fe (Möller and Bitzek, 2014). Previous studies also show that calculation of many properties concerning the Fe-H interaction by using this potential, such as the H dissolution energies, H diffusion barrier, binding of H to vacancy and free surfaces in α-Fe, diffusion barriers of hydrogen-vacancy pair, etc., can give very good agreement with the DFT results (Hayward and Fu, 2013; Ramasubramaniam et al., 2009). We can therefore consider this potential as rather reliable for the atomistic modeling of α-Fe with hydrogen.

Two types of individual GBs in α-Fe, the $\Sigma9\{1\,\overline{1}0\}$ pure twist GB (hereafter referred to as Σ9-GB) and $\Sigma31<3\,2\,0>\{\overline{14}\,21\,\overline{13}\}$ symmetric tilt GB (hereafter referred to as Σ31-GB), were selected for the study. The large bicrystal model, as illustrated in Fig. 1a-c, was employed for the MD simulation study of the individual GBs. The dimensions of the bicrystal models for the Σ9-GB (Fig. 1a) and Σ31-GB (Fig. 1b) are 17.2 nm $\times$ 21.8 nm $\times$ 32.4 nm and 22.6 nm $\times$ 16.6 nm $\times$ 32.5 nm, respectively. The models have 1,036,800 Fe atoms for the Σ9-GB and 1,026,240 Fe atoms for the Σ31-GB. For investigation of the hydrogen effect, models filled with H atoms, which correspond to an average hydrogen concentration of approximately 26, 66, 110, 187 mass ppm for the Σ9-GB, and 32, 64, 115, 139 mass ppm for the Σ31-GB, were prepared. Given that the overall size (grain size) of the MD bicrystal models is quite limited (i.e., nanometer scale) in our study, the average hydrogen concentration in the MD bicrystal models should be served as a reference variable. The values of the hydrogen concentrations in the models here appear to be high, but should still be comparable to those in the experiments (Nagumo, 2016; Tiegel et al., 2016) by further considering that the values of hydrogen contents in metals as reported by the experiments in many cases are likely to be those hydrogen atoms which are diffusible in the material. Because of the application of periodic boundary condition to all three dimensions, two identical GBs were included in the bicrystal model, as shown in Fig. 1a-c.

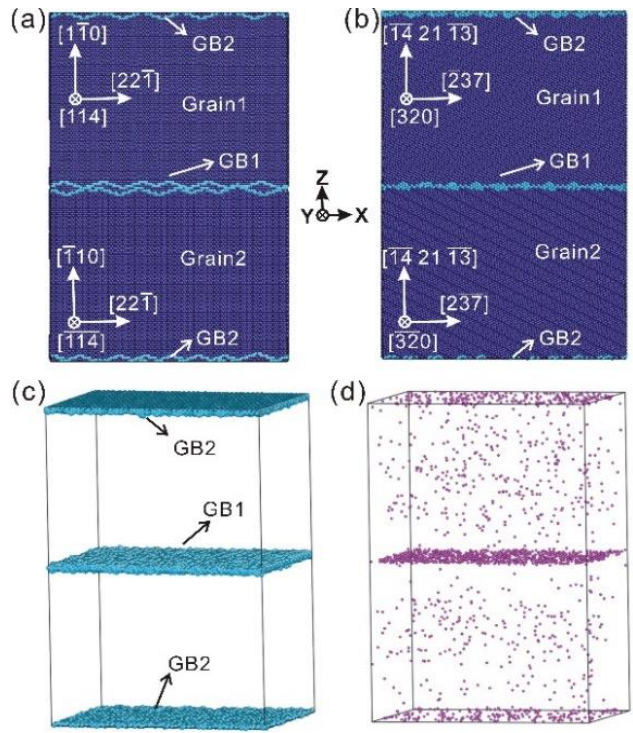

**Fig. 1.** Bicrystal models for MD tensile simulation. (a) Bicrystal model for the Σ9-GB in α-Fe without H atoms. (b)(c) Bicrystal model for the Σ31-GB in α-Fe without H atoms. (d) Distribution of H atoms in bicrystal model for the Σ31-GB in α-Fe with 32 mass ppm H atoms after MD heat treatment simulation. The dark blue spheres represent the Fe atoms in perfect α-Fe lattice. The light blue spheres represent the Fe atoms which belong to crystal defects. The pink tiny spheres are the H atoms.

Before the MD tensile simulation, the initially constructed models were first heated at 1000 K for 100 ps, then cooled to 300 K in 100 ps, followed by further equilibrium at 300 K for 1.2 ns. The purpose of this heat treatment procedure is to bring the structure of the GB and the distribution of H atoms close to equilibrium state. The MD tensile simulation was then performed at 300 K with a constant strain rate of $1 \times 10^8$ s$^{-1}$. An adaptive time step with a maximum displacement of any atom of 0.05 Å in one integration step was adopted for the MD heat treatment simulation of the bicrystal models. Constant time steps of 1.0 fs and 2.0 fs were used in the room temperature MD tensile simulation for models with and without H atoms, respectively. The Nose–Hoover thermostat (Hoover, 1985; Nosé, 1984) was used to control the temperature of the models during the MD simulations, while the Parrinello–Rahman barostat (Parrinello and Rahman, 1981) was used to control the stress state of the models. For MD tensile simulations, a uniaxial tensile load was applied with the lateral stresses of the model set to zero to allow the relaxation of the model in the lateral dimensions accordingly.

For production of all illustrations in the article and the supplementary movies, snapshots of the models were first fast quenched to remove the thermal noise on atoms positions of the models. The centro-symmetry parameter (CSP) (Kelchner et al., 1998) was used to identify the Fe atoms which belong to crystal defects (i.e., GBs, dislocations, point defects, etc.). In the CSP method, we considered atoms with CSP value larger than 6.0 as the ones that constitute the defects for the purpose of visualization. Meanwhile, a structural disorder parameter (SDP) was defined for each atom as the averaged CSP value of all the atoms within the sphere of radius of 3.487 Å centered at the specific atom. The SDP was used to give illustration of the disordering of the atomistic structure of GB. The coordination number of atoms was used to identify the atoms on surface of a void. On calculation of the coordination number of each atom, only Fe atoms within the cutoff radius of 3.487 Å were counted as neighboring atoms. With this cutoff choice, an Fe atom in body centered cubic lattice normally has a coordination number of 14.

The atomic stress as proposed by Cormier et al. (Cormier et al., 2001) was used for the analysis of the local stress distribution in the model. This definition of the atomic stress for a particular atom in the atomistic model can be expressed as

$$\sigma_{ij}(r_c) = \frac{1}{\Omega} \left[ \sum_{\alpha} \frac{p_i^{\alpha} p_j^{\alpha}}{m^{\alpha}} \Lambda^{\alpha}(r_c) - \frac{1}{2} \sum_{\alpha \neq \beta} \frac{\partial V}{\partial r^{\alpha\beta}} \frac{r_i^{\alpha\beta} r_j^{\alpha\beta}}{|\vec{r}^{\alpha\beta}|} l^{\alpha\beta}(r_c) \right]. \tag{1}$$

Here, the Roman subscripts $i,j$ denote tensorial components and the Greek superscripts

$\alpha,\beta$ denote atomic labels. For this definition of atomic stress, a sphere with cutoff radius $r_c$ centered at the specific atom is defined. $\Omega$ stands for the volume of the sphere. $p^{\alpha}$ is the momentum of atom $\alpha$ with mass $m^{\alpha}$. $\Lambda^{\alpha}(r_c)$ is unity if atom $\alpha$ lies within the sphere and is zero otherwise. $V$ is the interatomic potential of the atomistic model and $\vec{r}^{\alpha\beta} = \vec{r}^{\alpha} - \vec{r}^{\beta}$. $l^{\alpha\beta}(r_c)$ is the fraction of the length of the $\alpha-\beta$ bond that lies within the sphere ($0 \leq l^{\alpha\beta} \leq 1$). In our calculations of the atomic stress, a cutoff radius ($r_c$) of 5.5 Å was adopted for identification of neighboring atoms in the local sphere of each atom.

To measure the total volume of voids formed in the model on loading, a numerical Monte Carlo method was used and it proceeds as follows. For a given snapshot of the model captured in the MD tensile straining process, a large number ($N_{total}$) of geometry points in the simulation box of the model are selected randomly. For each point, a cubic volume of side length of 4.0 Å which centers on the point is defined. If there is no Fe atom within the cubic volume, we consider this point as belong to a void (the distance of two nearest Fe atoms in perfect BCC lattice is 2.5 Å). In this way, the number ($N_{voids}$) of those geometry points selected which belong to any of the voids formed in the model can be calculated. Provide $N_{total}$ is large enough ($N_{total} = 1 \times 10^6$ in our calculation), the value of $N_{voids}/N_{total}$ will converge to the fraction of $V_{voids}/V_{box}$ with marginal numerical error. Here $V_{voids}$ is the total volume of all the voids formed in the model, and $V_{box}$ is the volume of the simulation box which can be easily obtained. The total volume of all the voids formed in the model ($V_{voids}$) can then be calculated straightforwardly.

## 2.2 GB cohesive strength calculation

For a quantitative characterization of the cohesive strength of GBs, molecular statics tensile simulation of small bicrystal models for the GBs was performed by using the same EAM interatomic potential for Fe-H. The bicrystal models used in this calculation were made as small as possible to suppress the dislocation activities which would otherwise interfere with the GB cleavage process. Here we have performed the GB cohesive strength calculation for both the $\Sigma$9-GB and $\Sigma$31-GB in $\alpha$-Fe (with and without H atoms). The molecular statics tensile loading was performed by stretching the model along the axial perpendicular to the GB plane in 0.3 Å steps to fracture. Two border slabs were devised in the model to serve as the 'grips' on stretching, while periodic boundary condition was applied on the dimensions parallel to the GB. At each stretching step, a relaxation of the atoms in the interior of the model by the conjugate gradient energy minimization method was performed while keeping the atoms in the border slabs fixed. The energy minimization stops when no force on any atom in the interior of the model is larger than $10^{-8}$ eV/ Å. During the molecular statics loading procedure, the size of the lateral dimensions of the model was fixed. The maximum stress on the stress-displacement curve was defined as the cohesive strength of the GB.

For comparison, we have also performed the GB cohesive strength calculation using the first-principles DFT method for the Σ9-GB. The DFT calculation was performed by using the VASP code (Kresse and Furthmüller, 1996). The electron-ion interaction was described using projector augmented wave (PAW) method (Kresse and Joubert, 1999). The exchange correlation between electrons was treated with generalized gradient approximation (GGA) in the Perdew-Burke-Ernzerhof (PBE) form (Perdew et al., 1996). We used an energy cutoff of 268 eV for the plane wave basis set for all supercells. The Brillouin-zone integration was performed within the Monkhorst-Pack scheme using k meshes of $(3 \times 4 \times 1)$ for all supercells. The stopping criteria for the electronic structure optimization was set to $10^{-4}$ eV. The molecular statics loading simulation by using the DFT method was performed in the same way as the simulation using the EAM potential, except that the ionic relaxation in energy minimization stops when the change of total energy between successive iterations is smaller than $10^{-4}$ eV in the DFT calculation. Since atoms in the border slabs were fully relaxed before stretching and were frozen during ionic relaxation of the model on stretching, the stress obtained as calculated by the Hellman-Feynman force theorem for the bicrystal model as a whole was then divided by the volume fraction of the interior of the model.

## 3. Results

### 3.1 Tensile simulation of large bicrystal models

As seen in Fig. 1d, the H atoms will segregate to the GB because of the strong H trapping effect of the GB at room temperature and the high diffusivity of H atoms in the α-Fe lattice. Table 1 summaries the volume density of H atoms in the GB core and grain interior for the models filled with hydrogen. Fig. 2a and b show the stress–strain curves for the MD tensile simulation of models with the two types of GBs, respectively. For all the models, the plastic deformation is realized by dislocation emission from GB, and these dislocations will eventually impinge on the other GB in the model (see Fig. 2c and Supplementary Movies M1, M2).

**Table 1**

Volume density of H atoms in the core of GB1 and GB2 ($\rho_{H[GB1]}$, $\rho_{H[GB2]}$), as well as in the grain interior of Grain1 and Grain2 ($\rho_{H[Grain1]}$, $\rho_{H[Grain2]}$) for the bicrystal models (see Fig. 1). For calculation of the volume density of H atoms, a slab with thickness of 1.2 nm was used to define the region of GB core, and a slab with thickness of 9.0 nm in the center of the grain was used to define the region of grain interior. The unit for the H atoms volume density is nm$^{-3}$ (1 nm$^{-3}$ ≈ 214 mass ppm for H in α-Fe lattice). The bicrystal models are distinguished by the GB type together with the concentration of H atoms in the models correspondingly.

| Model | $\rho_{H[GB1]}$ | $\rho_{H[GB2]}$ | $\rho_{H[Grain1]}$ | $\rho_{H[Grain2]}$ |
|---|---|---|---|---|
| Σ9-GB 26 mass ppm | 1.541 | 0.369 | 0.076 | 0.074 |
| Σ9-GB 66 mass ppm | 3.805 | 1.213 | 0.176 | 0.173 |
| Σ9-GB 110 mass ppm | 6.446 | 1.766 | 0.309 | 0.306 |
| Σ9-GB 187 mass ppm | 11.542 | 3.309 | 0.461 | 0.495 |
| Σ31-GB 32 mass ppm | 1.794 | 1.029 | 0.076 | 0.058 |
| Σ31-GB 64 mass ppm | 3.204 | 2.056 | 0.169 | 0.135 |
| Σ31-GB 115 mass ppm | 5.510 | 3.541 | 0.321 | 0.289 |
| Σ31-GB 139 mass ppm | 5.951 | 4.439 | 0.436 | 0.342 |

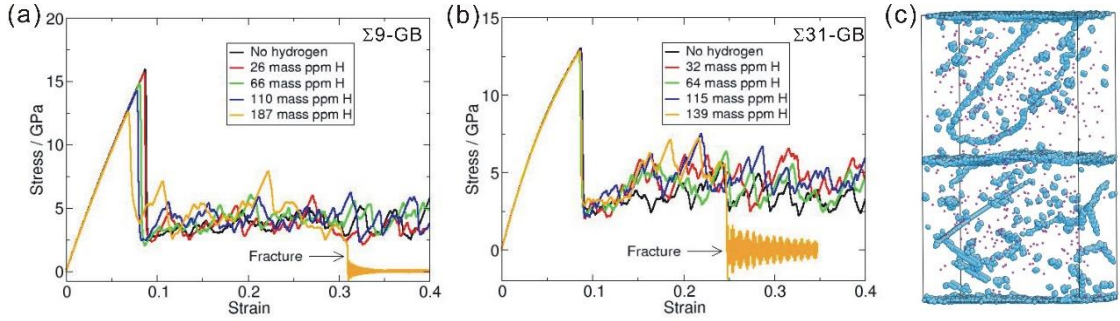

**Fig. 2.** (a)(b) The stress–strain curves for MD tensile simulation of the Σ9-GB and Σ31-GB, respectively. (c) A snapshot of the model for the Σ31-GB with 32 mass ppm H atoms at the tensile strain of 0.108. The Fe atoms in perfect α-Fe lattice are not shown.

One can notice from the curves in Fig. 2a that the initial yield stress of the bicrystal models for the Σ9-GB decreases with increasing H concentration. Because there is no pre-existing dislocation in the models, this indicates that H can facilitate dislocation emission from the GB. On the other hand, the nearly identical initial yield stress of the bicrystal models for the Σ31-GB as shown in Fig. 2b suggests that hydrogen has limited effect on dislocation emission from this type of GB. For dislocation emission from GB, structural rearrangement of atoms in the GB zone intersecting the glide plane of the dislocation emitted is normally required. Such kind of structural rearrangement of atoms often involves some atomic shuffling motions (Du et al., 2016). With hydrogen atoms segregated on the GB, the structural rearrangement and the shuffling motions of atoms in the GB zone can be promoted (like the Σ9-GB) or hindered or not affected (like the Σ31-GB), depending on the specific type of GB concerned. A comprehensive understanding of the hydrogen effect on the behavior of dislocation emission from GB demands investigation of many other types of GBs, which is beyond the scope of this article.

### 3.2 H promotes void formation on GB

Besides the dislocation activities, structural analysis of the models indicates that voids can form on the GB during tensile loading. Fig. 3 shows the atomic configurations of the models at the end of the MD tensile simulations (tensile strain of 0.400) for the Σ9-GB. With only the atoms of coordination number less than 10 in the models visualized (coordination number of an atom in perfect α-Fe lattice is 14), the voids can be identified. For the models without H atoms (Fig. 3a) and with 26 mass ppm H atoms (Fig. 3b), only very small embryos of voids containing just a few vacancies are visible. With higher hydrogen concentrations in the model, voids of significant size can form at this stage of tensile straining (Fig. 3c and d). The small voids coalesce into larger ones, as seen in Fig. 3d. If the hydrogen concentration is sufficiently high (Fig. 3e), the model can even fracture in a cleavage style via void extension along the GB in early stage of plastic straining of the model (see the corresponding stress–strain curve in Fig. 2a). A vivid demonstration of voids nucleation and growth, and also the fracture processes, can be found in the Supplementary Movies M3 and M4. The similar behavior of hydrogen concentration dependent void formation on GBs can be recognized in the MD tensile simulation of the Σ31-GB as well (see Fig. 4 and Supplementary Movies M5 and M6).

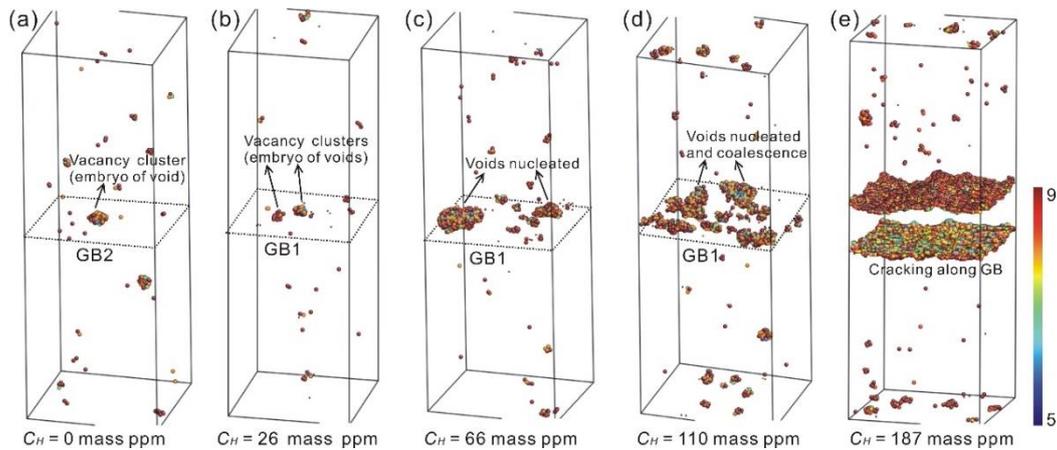

**Fig. 3.** (a)-(e) Snapshots of bicrystal models for the Σ9-GB with 0, 26, 66, 110, and 187 mass ppm H atoms, respectively. All snapshots were captured at the end of MD tensile straining simulation (tensile strain of 0.400). Atoms are colored based on their coordination number. Only atoms with coordination numbers less than 10 are displayed. The dotted frames indicate the position of GBs in the bicrystal models.

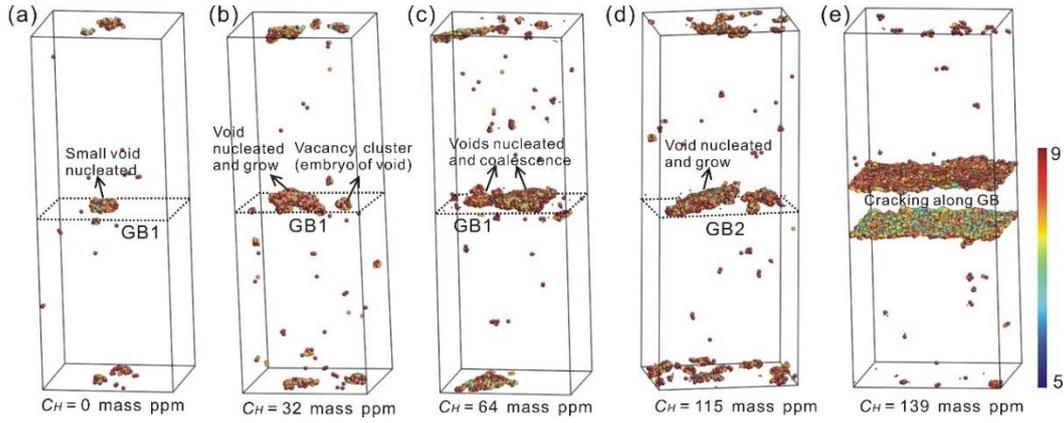

**Fig. 4.** (a)-(e) Snapshots of bicrystal models for the Σ31-GB with 0, 32, 64, 115, and 139 mass ppm H atoms, respectively. All snapshots were captured at the end of MD tensile straining simulation (tensile strain of 0.400). Atoms are colored based on their coordination number. Only atoms with coordination numbers less than 10 are displayed. The dotted frames indicate the position of GBs in the bicrystal models.

For each model, the total volume of all voids formed during the tensile straining can be roughly measured. Fig. 5a and b show the curves for variation of the total volume of voids with strain in the models with the Σ9-GB and the Σ31-GB, respectively. Again, the enhancing effect of hydrogen on void nucleation and growth on tensile loading of the models is evident. The hydrogen concentration dependent void formation behavior on tensile straining is clear evidence of the embrittling effect of hydrogen in α-Fe.

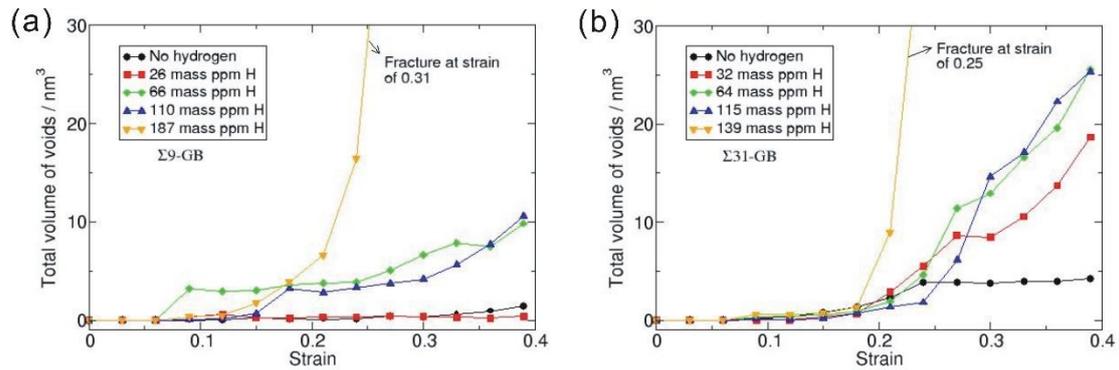

**Fig. 5.** Evolution of the total volume of all voids formed during tensile straining of the bicrystal models for the Σ9-GB (a) and the Σ31-GB (b).

3.3 GB decohesion aided by dislocation-GB reaction

A close examination of the structural evolution of the models reveals that, before the void nucleation/growth and the cracking processes, a significant amount of dislocation impingement and emission events can occur in the specific GB area where the void nucleates and grows (see Supplementary Movies M3-M6). It is therefore very likely that the dislocation-GB reaction by dislocation impingement/emission on GB plays an important role in void nucleation and growth on GB, or even dictates decohesion of the GB for the void nucleation and growth to proceed. To verify this assumption, the dislocation-GB reaction process was then carefully examined.

Fig. 6a and b show one cross section around the GB in the model with 66 mass ppm H atoms for the Σ9-GB before tensile loading and at a tensile strain of 0.200, respectively. Before tensile loading, the atomistic structure of this Σ9{1 $\bar{1}$0 } pure twist GB in α-Fe is fairly ordered (Fig. 6a and c). After a certain amount of dislocations impingement and emission at the GB on loading to tensile strain of 0.200, the local atomistic structure of the GB has changed at many sites on the GB, as marked by the arrows in Fig. 6b. In the enlarged view, a comparison of Fig. 6c and Fig. 6d illustrates that the local atomistic structure of the GB becomes much more disordered because of the dislocation-GB reaction at these sites.

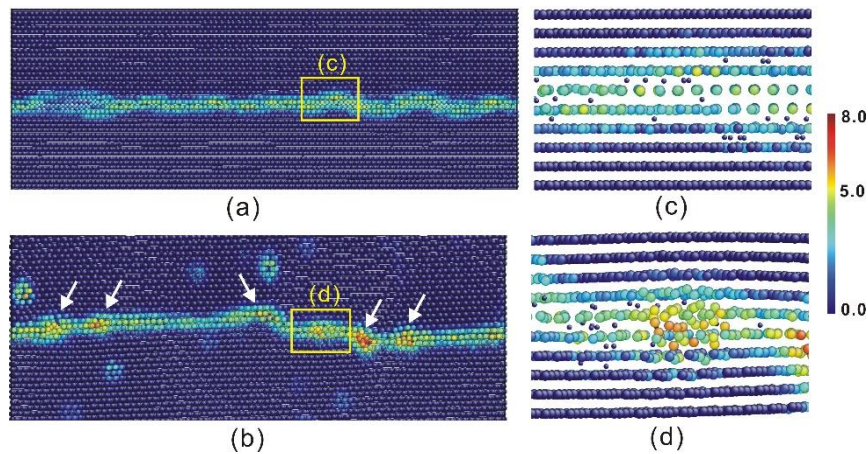

**Fig. 6.** Local views of a slice of thickness 1.9 nm taken around the GB from the model of the Σ9-GB with 66 mass ppm H atoms before loading (a) and at the tensile strain of 0.200 after a certain amount of dislocations emitted/impinged on the GB (b). (c) and (d) are the enlarged view of the area as indicated by the rectangles in (a) and (b), respectively. The coloring of the Fe atoms is based on their SDP value (see the color

bar on the right hand side). The smaller dark blue spheres are the H atoms.

Fig. 7 shows another cross section around the GB in the model with 66 mass ppm H atoms for the Σ9-GB at four different strains (i.e., 0.000, 0.260, 0.280 and 0.330) during the MD tensile simulation. The top panels (Fig. 7a-d) show that, with dislocations impingement and emission on the GB, the local atomistic structure of the GB will significantly change. The middle panels (Fig. 7e-h) suggest that, the void (marked by dotted circle) nucleates on the GB only after a certain amount of dislocations impinged/emitted on the specific GB area. Meanwhile, Fig. 7j shows that, in addition to the change of the local atomistic structure on the GB, the dislocation impingement on the GB can also result in local dilative stress concentration at the reaction site on the GB (see the area indicated by the arrow in Fig. 7j). The local stress concentration can be relaxed to a large extent with void nucleation there (Fig. 7k). Further straining will lead to void growth (Fig. 7h) and a redistribution of the local stress around the void (Fig. 7l).

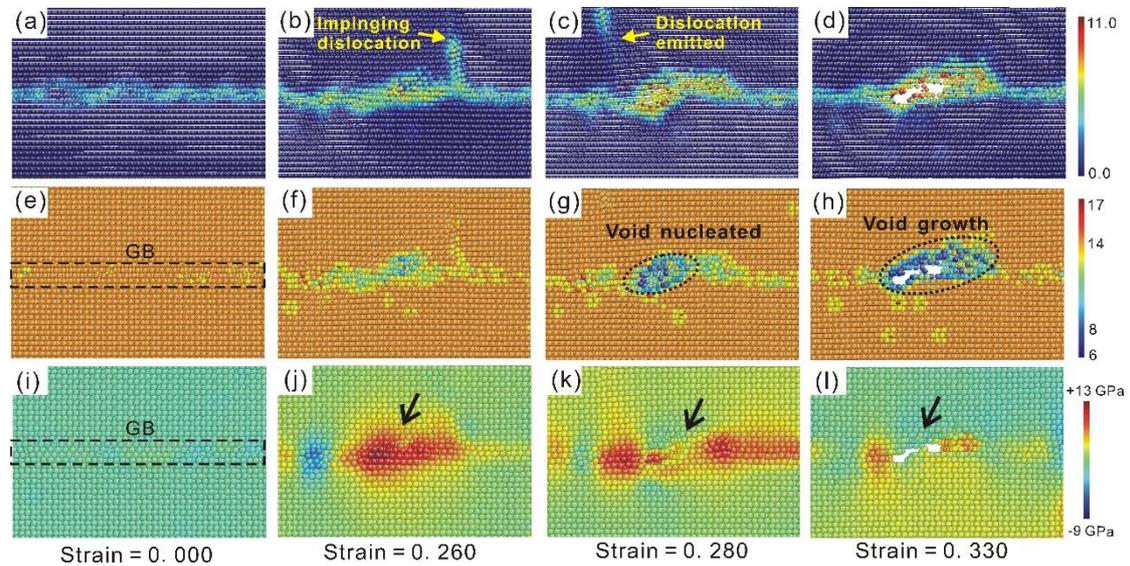

**Fig. 7.** Local views of a slice of thickness 1.9 nm taken around the GB from the model of the Σ9-GB with 66 mass ppm H atoms at tensile strains of 0.000 ((a)(e)(i)), 0.260 ((b)(f)(j)), 0.280 ((c)(g)(k)) and 0.330 ((d)(h)(l)). For the three views of each strain value, the atoms are colored based on their SDP value (top panels), coordination number (middle panels), and local triaxial stress (bottom panels). Atoms of coordination number less than 11 are considered to be surface atoms of the void, as marked by the dotted circles in (g) and (h). In (e) and (i), the dashed rectangles indicate the GB.

Similar behavior of dislocation-GB reaction can be identified for other void nucleation processes as well, as can be seen in Fig. 8 and Fig. 9 for another example in MD tensile simulation of the model with the Σ31-GB and with 32 mass ppm H atoms. With the above results, the fracture mechanism and the H embrittling effect can be envisioned as follows. For models filled with hydrogen, the GBs will be largely segregated with H atoms. By applying a tensile load, dislocation plasticity with dislocation emission/impingement on the GBs can be triggered. The role played by the dislocation plasticity is two-fold. Firstly, the local atomistic structure of GBs can be changed and become more disordered because of dislocation impingement/emission on GBs. The disordered atomistic structure of the GB can be viewed as an 'activated state' of the GB (e.g., Fig. 6d and Fig. 8d), as compared with the original 'ground state' of the GB before loading (e.g., Fig. 6c and Fig. 8c). Secondly, the dislocations impinged/emitted on the GB can result in a local dilative stress concentration at the reaction sites on GB. With these effects combined, debonding of atoms at these sites on GB can be greatly facilitated, and a void would then nucleate and grow on the GB. It is quite likely that the embrittling effect of H atoms can be largely attributed to the activation of GB for which the cohesive strength of the 'activated state' of GB is significantly reduced with the presence of H atoms.

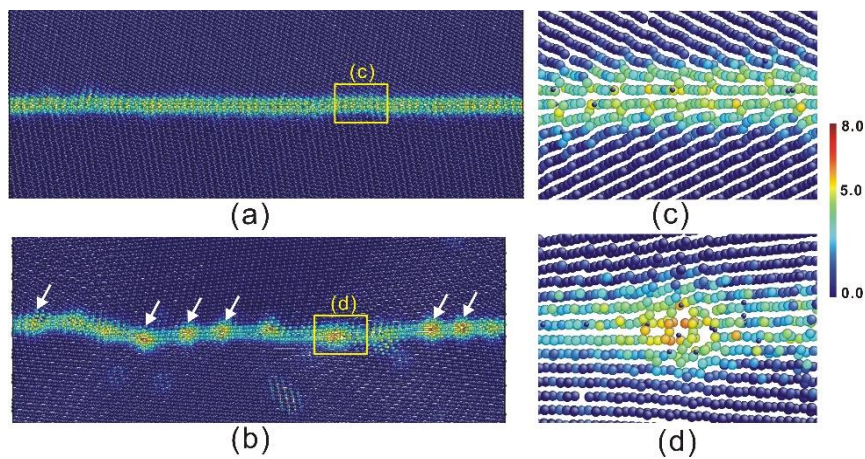

**Fig. 8.** Local views of a slice of thickness 2.3 nm taken around the GB from the model of the Σ31-GB with 32 mass ppm H atoms before loading (a) and at the tensile strain of 0.200 after a certain amount of dislocations emitted/impinged on the GB (b). (c) and (d) are the enlarged view of the area indicated by the rectangles in (a) and (b), respectively. The coloring of the Fe atoms is based on their SDP value (see the color bar on the right hand side). The smaller dark blue spheres are the H atoms.

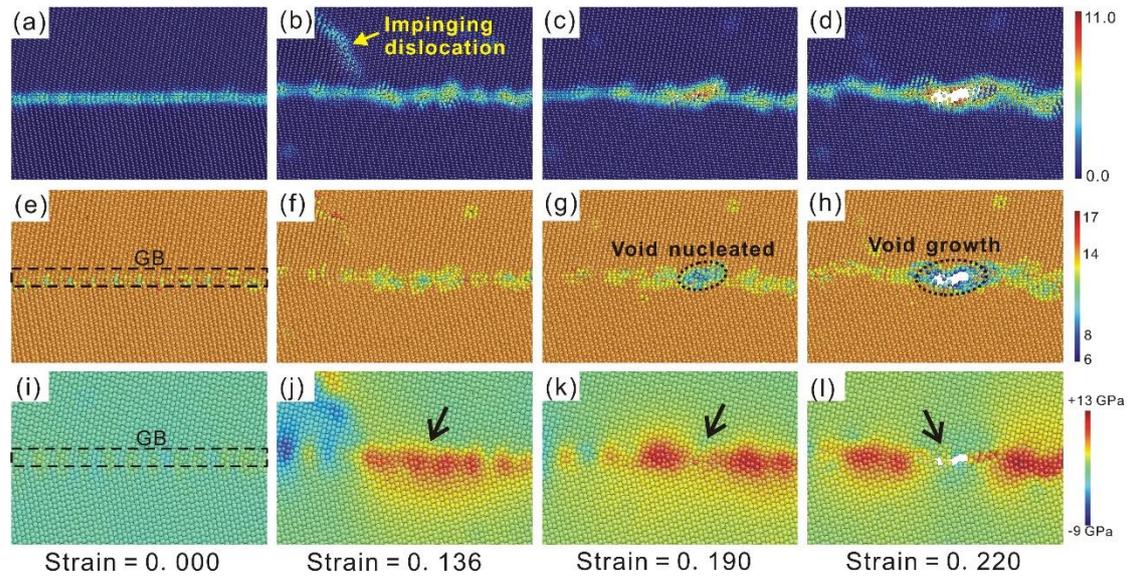

**Fig. 9.** Local views of a slice of thickness 2.3 nm taken around the GB from the model of the Σ31-GB with 32 mass ppm H atoms at tensile strains of 0.000 ((a)(e)(i)), 0.136 ((b)(f)(j)), 0.190 ((c)(g)(k)) and 0.220 ((d)(h)(l)). For the three views of each strain value, the atoms are colored based on their SDP value (top panels), coordination number (middle panels), and local triaxial stress (bottom panels). Atoms of coordination number less than 11 are considered to be surface atoms of the void, as marked by the dotted circles in (g) and (h). In (e) and (i), the dashed rectangles indicate the GB.

3.4 Mechanical effect of GB activation

　　The cohesion of a GB segregated with H atoms can be severely weakened by the activation of the GB to a more disordered atomistic structure due to dislocation-GB reaction, as can be further demonstrated via a calculation of the cohesive strength of the GB before and after dislocation-GB reaction. Fig. 10 shows the small bicrystal models for the Σ9-GB in α-Fe with different atomistic structures. The models contain zero (Fig. 10a and b), or four H atoms (Fig. 10c and d) for which the atom density of hydrogen in the GB zone is comparable to the above large model with 110 mass ppm hydrogen concentration. In Fig. 10a and c, the structure of the GBs was obtained by a near global minimization procedure and can be considered as the 'ground state' of the GB before dislocation-GB reaction. The structure of the GBs in Fig. 10b and d was obtained by a random displacement of atoms in the GB zone of the models in Fig. 10a and c, following by a local minimization procedure, respectively. It can be seen that both the GBs in Fig. 10b and d are trapped in an 'activated state' with a more disordered atomistic structure as compared with those in Fig. 10a and c, respectively. The artificially disturbed atomistic structure introduced on the GBs in Fig. 10b and d can serve as a good imitation of the activated structure of GB with similar disorder as yielded by dislocation-GB reaction (see Fig. 6d for example).

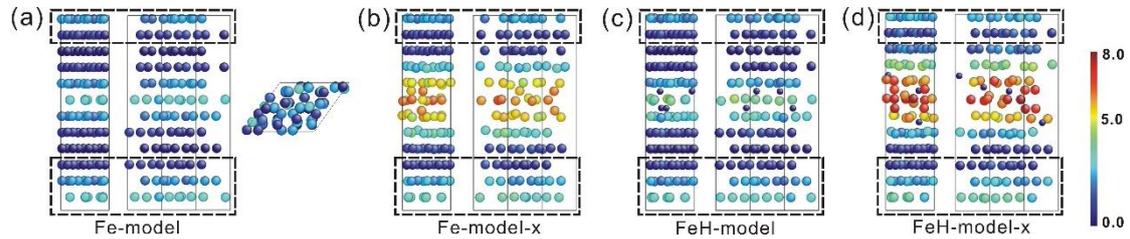

**Fig. 10.** Small bicrystal models of the Σ9-GB in α-Fe with different GB atomistic structures to model the local GB area before and after dislocation-GB reaction. (a)(b) Bicrystal model for the GB without H atoms. (c)(d) Bicrystal models for the GBs with four H atoms in the GB core. For each model, two orthogonal views of the model are displayed. A projected view of the model along the normal of GB is given in (a) as well. The coloring of the Fe atoms is based on their SDP values (see the color bar on the right hand side of the figure). The smaller dark blue spheres are the H atoms. The dashed frames mark the border slabs of the bicrystal model.

Fig. 11a shows the stress–displacement curves for the molecular statics tensile stretching of the models in Fig. 10 by using the EAM potential for Fe-H. The cohesive strength of the corresponding GBs can be obtained as the peak stress on these stress-displacement curves. Since there is randomness in both the distribution of H atoms in GB zone and the disordered structure of the 'activated state' artificially introduced on GBs in the small bicrystal models, we have prepared a number of models with different distribution of H atoms in the GB zone, as well as a number of models with different disordered atomistic structure of the 'activated state' of the GB. The statistically averaged results of the GB cohesive strength calculated from all of these models are given in Table 2. As can be seen from the curves in Fig. 11a and the values in Table 2, for GBs in the 'ground states' (e.g., Fig. 10a and c), the introduction of H atoms can only decrease the cohesive strength of the GB very slightly, i.e., a reduction of 1.56 GPa for the Σ9-GB and 0.68 GPa for the Σ31-GB. By comparison, a reduction in the GB cohesive strength of approximately 4.48 GPa for the Σ9-GB and 2.02 GPa for the Σ31-GB can result simply from the activation of the GB segregated with H atoms (e.g., from the GB in Fig. 10c to that in Fig. 10d).

Similar calculation by using the first-principles DFT method for the same models with the Σ9-GB as shown in Fig. 10 has also been performed. Fig. 11b shows the stress-displacement curves for molecular statics stretching of these models by using the DFT method. A comparison of the curves in Fig. 11b with those in Fig. 11a indicates that the GB cohesive strength calculated by these two methods agrees reasonably well with each other, considering that a rougher stopping criteria for the ionic relaxation by conjugate gradient energy minimization has been adopted in the DFT calculation. It is clear that as a result of dislocation-GB reaction, the activation of the GB segregated with H atoms can greatly facilitate decohesion of the GB.

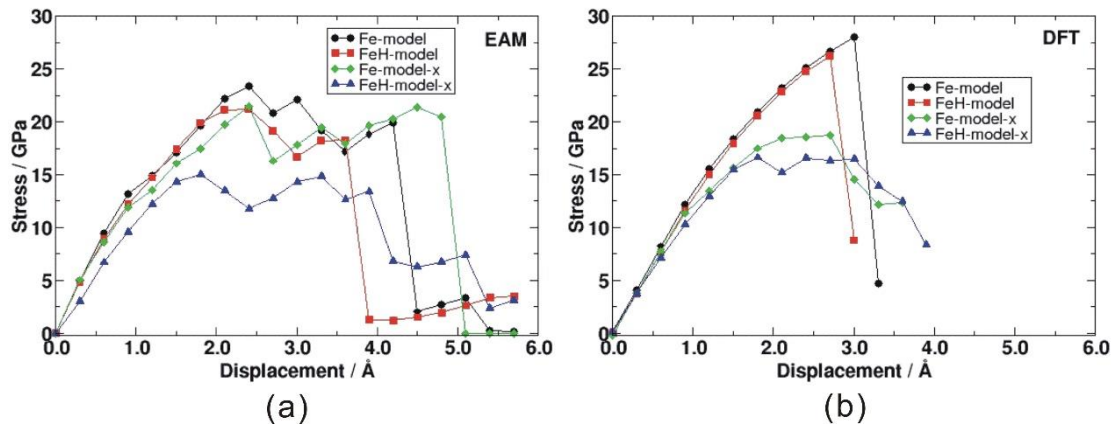

(a)                                           (b)

**Fig. 11.** The stress–displacement curves for molecular statics tensile stretching of the small bicrystal models in Fig. 10 by using both the EAM potential for Fe-H (a) and the first-principles DFT method (b).

**Table 2**

The GB cohesive strength of the $\Sigma$9-GB and $\Sigma$31-GB in $\alpha$-Fe calculated by using the EAM potential for Fe-H.

| GB state | GB cohesive strength (GPa) | |
|---|---|---|
| | $\Sigma$9-GB | $\Sigma$31-GB |
| without H 'Ground state' | 23.25 | 19.92 |
| with H [a] 'Ground state' | 21.69 $\pm$0.56 [b] | 19.24 $\pm$0.84 [b] |
| without H 'Activated state' | 19.74 $\pm$1.33 [c] | 19.18 $\pm$0.40 [c] |
| with H [a] 'Activated state' | 17.21 $\pm$1.98 [c] | 17.22 $\pm$0.94 [c] |

[a] The area density of H atoms in the GB core is 7.72 $nm^{-2}$ and 6.93 $nm^{-2}$ for the models with $\Sigma$9-GB and $\Sigma$31-GB, respectively.

[b] Five models with different distribution of H atoms in GB zone were used in calculation. The averaged value of the GB cohesive strength calculated from the five models and the standard deviation of the values are given in the table accordingly.

[c] Five models with different disordered atomistic structure of GB, as produced by varying the random number for preparation of the GB in 'activated state' through random displacement of atoms in GB zone, were used in calculation. The averaged value of the GB cohesive strength calculated from the five models and the standard deviation of the values are given in the table accordingly.

# 4. Discussion

The two types of GBs (the $\Sigma$9-GB and $\Sigma$31-GB in $\alpha$-Fe) selected in our investigation can be viewed as rather representative of the general high angle GBs in real polycrystals of $\alpha$-Fe. Based on the above results, we hereby propose that the dislocation-GB reaction via dislocation impingement/emission on the GB should play a key role in hydrogen induced fracture of polycrystalline α-Fe and other metals. As schematically illustrated in Fig. 12, in the scenario that the dislocations impingement/emission on GB serving as a determining step for void nucleation and growth, the characteristics of experimentally observed fractograph and microstructure associated with hydrogen induced fracture can be satisfactorily accounted for as follows. If the hydrogen concentration is not high (Fig. 12a), void nucleation and its growth on the GB would require significant amount of dislocation activities nearby the GB so as to give the necessary amount of dislocation-GB reaction to aid local decohesion of the GB. Eventually, fracture of the material will yield a quasi-cleavage fracture surface with features like tears, shallow dimples, etc. due to the local plastic strain nearby the cracking GBs. And the traces of the original GBs will be hardly recognized on the fractograph. It is interesting to notice that, some quasi-cleavage fracture surfaces observed in the hydrogen embrittlement experiments can indeed be identified as traces of GB planes of the materials before fracture (Nagao et al., 2012; Shibata et al., 2015; Shibata et al., 2012). We suggest that many more cases of such kind of correspondence on fractograph should be able to be identified in the experimental observations with a careful analysis.

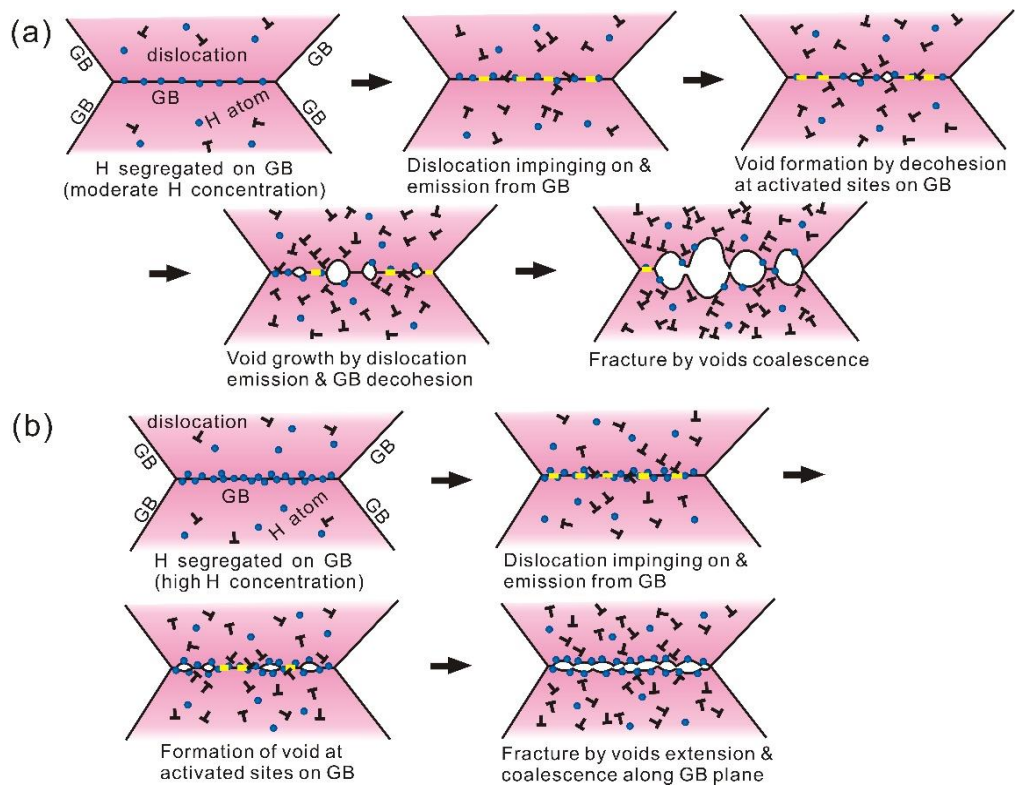

**Fig. 12.** Schematic illustration of H embrittlement controlled by dislocation-GB reaction in metals. (a) Polycrystalline sample with moderate concentration of hydrogen mechanically failed with a quasi-cleavage fracture surface on which the traces of the original GBs can hardly be identified. (b) Intergranular fracture of a polycrystalline sample with fairly high hydrogen concentration for which the traces of original GBs are largely preserved on the fracture surface. The yellow strips on the GBs represent sites at which the reaction of GB with dislocations results in a locally activated state of the GB.

At higher hydrogen concentrations (Fig. 12b), less dislocation activities nearby the GB are required to trigger local decohesion of the GBs on cracking through the dislocation-GB reaction. In such case, the traces of the original GBs can be largely preserved on the fracture surface, and the fracture surface will be more intergranular. As revealed by the experiment, there can be arrays of dislocation slip traces on the hydrogen induced intergranular fracture surfaces (Martin et al., 2012). As long as the hydrogen concentration is not exceedingly high, dislocation activities nearby the GBs is normally required for the fracture to proceed. The dislocation activities around GBs on cracking will eventually yield a well-evolved dislocation structure in the volume of material just beneath the fracture surface for both the quasi-cleavage and intergranular style fracture as observed in many experiments (Martin et al., 2011a; Martin et al., 2011b; Martin et al., 2012; Nagao et al., 2012; Neeraj et al., 2012; Robertson et al., 2012; Robertson et al., 2015; Wang et al., 2014).

It needs to be mentioned that, the bicrystal models used here for MD tensile simulations are relatively small. There is no pre-existing dislocation in the model, and the strain rate is rather high in the MD tensile simulations. In practice, polycrystalline samples can have a certain amount of pre-existing lattice dislocation sources. Dislocation pile-ups at GBs can normally form on plastic straining, which will introduce local stress concentration on GBs. Also the strain rate is much lower than was assumed in our simulation. Thus, the yield stress and flow stress will be significantly lower in a realistic situation for the impingement and emission of dislocations on GBs.

However, the characteristics of dislocation-GB reaction by the impingement of dislocations on GB and subsequent emission of dislocations from the GB should be similar to those described here.

Previously, the dislocation-GB interaction has already been proposed to be able to contribute to the hydrogen induced intergranular fracture through the following three effects (Robertson et al., 2015; Wang et al., 2014): (i) H atoms can be transported along with the impinging dislocation to GB; (ii) an increase in strain energy density within the GB can be induced on the sites where lattice dislocations are incorporated into the GB; (iii) a local stress concentration can be established by the dislocation interaction with the GB. Our modeling results presented here clearly support the third point, as can be seen in the Figs. 7 and 9. For the increase in strain energy density within GB, our results as demonstrated in Figs. 6 and 8 indicate that it should correspond to the local activation of the GB to a more disordered atomistic structure on the atomic-scale level. As for transportation of H atoms to GB by gliding dislocations, a rough measurement shows that there is no significant increase of the amount of H atoms on the GB for all the cases studied in our MD tensile simulation. Here we suggest that the realization of hydrogen embrittling effect in polycrystalline materials is largely assisted by the activation of the GB to a more disordered atomistic structure through the dislocation-GB reaction. And we emphasize that the dislocation-GB reaction not only drives the hydrogen induced intergranular fracture, but also serves as a key process in the hydrogen induced quasi-cleavage fracture, as illustrated in Fig. 12.

As it may appear rather trivial that the activation of the GB to a more disordered atomistic structure with relatively higher GB energy will normally reduce the strength of the GB even without H atoms (Tucker and McDowell, 2011), we would like to suggest that the presence of H atoms on the GB can help stabilize the activated structure during the dislocation-GB reaction. As can be seen from Fig. 11a and Table 2, for GBs without H atoms, the reduction of the GB cohesive strength by activation of the GB is limited, which can be attributed to the fact that the GB is more prone to be recovered from the activated state during ionic relaxation without the presence of H atoms in GB zone. On the other hand, our calculation reveals that the introduction of H atoms on the GB can slightly decrease the GB cohesive strength without the activation of the GB (see Fig. 11 and Table 2), which is in good agreement with previous calculations for H segregation on GBs (Geng et al., 1999; Huang et al., 2017; Solanki et al., 2012; Tahir et al., 2014; Wang et al., 2016). While this chemical effect of bond weakening by hydrogen has long been recognized as the root of hydrogen embrittlement as proposed by the HEDE theory (Gerberich, 2012; Oriani, 1987; Troiano, 1960), we suggest that the GB activation by dislocation-GB reaction is at least equally important in realization of the hydrogen embrittling effect in polycrystalline materials. In this sense, the hydrogen embrittlement model as depicted in Fig. 12 can be viewed as an improved HEDE theory.

Nevertheless, one still needs to be reminded that, it is less likely that any single mechanism for hydrogen embrittlement is exclusively responsible for all hydrogen induced embrittlement process. Firstly, there are experimental observations that the hydrogen assisted cracking sometimes initiates or propagates at the intersection of localized slip bands in the grain interior in some metals (Nibur et al., 2009; Tarzimoghadam et al., 2017; Zhang et al., 2016). For such cases, it is likely that the dislocation-dislocation reaction can result in an 'activated structure' in the highly tangled dislocation junctions. The H atoms present in the 'activated structure' can greatly facilitate void formation there, as can be envisioned as a generalization of the dislocation-GB reaction mediated hydrogen embrittlement picture described here. However, further studies are required to confirm this point.

Secondly, the hydrogen embrittlement model proposed here does not exclude the importance of hydrogen transportation and gathering in the material on loading for the occurrence of hydrogen embrittlement. Our modeling results (Figs. 3-5) indicate that the nucleation and growth of voids on loading can be accelerated by an increase of the coverage of H atoms on the GB. This agrees with the experimental result that suppression of hydrogen embrittlement can be induced by a refinement of the grain size of the material for which the coverage of H atoms on GB is reduced under the same hydrogen charging condition (Bai et al., 2016). For the gathering of H atoms in the material, one particular case is the gathering of H atoms at the crack tip due to the stress concentration at the mode I crack loading (Song and Curtin, 2013). In the hydrogen embrittlement model proposed here, the crack corresponds to voids formed and coalescence on the GB (Fig. 12). Therefore, the grain size can be an important parameter for determination of the GB crack tip stress field, which was adopted as a priori assumption in a previous rather successful analysis of hydrogen embrittlement phenomena (Song and Curtin, 2013). This success, in return, lends credence to the validity of the hydrogen embrittlement model proposed here.

With the hydrogen embrittlement model proposed here, a theoretical formulation which relates the voids nucleation/growth to the external stress, strain rate, temperature, dynamic hydrogen coverage on GB, microstructure characteristics such as grain size, distribution of precipitates, dynamic dislocation structure, etc., can be plausibly developed for a comprehensive description of the hydrogen embrittlement effect of polycrystalline metals. In addition, the alloying effect on enhancing or suppressing the hydrogen embrittlement of metals can also be investigated based on this hydrogen embrittlement model. For example, carbon atoms in iron and steels can exist in the Fe lattice as interstitial solutes, precipitate as carbides, segregate to dislocation in the form of Cottrell atmospheres, or segregate on the GBs. The cohesive strength of GBs as well as the dislocation-GB reaction can be changed by the carbon atoms segregated on the GBs with or without H atoms. The dislocation motions and the evolution of dislocation substructures at a larger scale can be altered by the carbon interstitials, carbide precipitates and the Cottrell atmospheres, while the occurrence and frequency of dislocation-GB reaction can therefore be affected. The carbon effect in hydrogen embrittlement of the ferrous alloys can then be examined and analyzed accordingly.

## 5. Conclusions

In conclusion, by atomistic modeling of the tensile response of individual GBs in a bicrystal model of α-Fe with different concentrations of H atoms, we have demonstrated that the dislocation-GB reaction by dislocation impingement/emission on the GB plays a key role wherein the hydrogen can degrade the mechanical performance of polycrystalline metals. The dislocation-GB reaction can result in a locally activated state of the GB with a more disordered atomistic structure of the GB, and also introduce a local dilative stress concentration there. For GB segregated with H atoms, the cohesion of the GB will be significantly weakened when it is in the activated state. The GB activation behavior may signify an important GB characteristic that has long been overlooked, and it should have important implications for design of new structural metals with improved resistance to hydrogen embrittlement.

**Acknowledgements**

This work was supported by NEDO Basic Research Program entitled "Technological Development of Innovative New Structural Materials", Grants-in-Aid for Scientific Research in Innovative Area (Grant No. 22102003), Scientific Research (A) (Grant No. 23246025), Challenging Exploratory Research (Grant No. 25630013), Elements Strategy Initiative for Structural Materials (ESISM), and the National Science Foundation of China (Grant No. U1760203).

**Supplementary Material**

Supplementary movies related to this article can be found at https://doi.org/10.1016/j.ijplas.2018.08.013.

# References

Ackland, G.J., Mendelev, M.I., Srolovitz, D.J., Han, S., Barashev, A.V., 2004. Development of an interatomic potential for phosphorus impurities in $\alpha$-iron. Journal of Physics: Condensed Matter 16, S2629-S2642.

Aoki, M., Saito, H., Mori, M., Ishida, Y., Nagumo, M., 1994. Deformation microstructures of a low carbon steel characterized by tritium autoradiography and thermal desorption spectroscopy. Journal of the Japan Institute of Metals-Nihon Kinzoku Gakkaishi 58, 1141-1148.

Bai, Y., Momotani, Y., Chen, M.C., Shibata, A., Tsuji, N., 2016. Effect of grain refinement on hydrogen embrittlement behaviors of high-Mn TWIP steel. Materials Science and Engineering: A 651, 935-944.

Birnbaum, H.K., Sofronis, P., 1994. Hydrogen-Enhanced Localized Plasticity - a Mechanism for Hydrogen-Related Fracture. Mat Sci Eng a-Struct 176, 191-202.

Cormier, J., Rickman, J.M., Delph, T.J., 2001. Stress calculation in atomistic simulations of perfect and imperfect solids. Journal of Applied Physics 89, 99.

Du, J.-P., Wang, Y.-J., Lo, Y.-C., Wan, L., Ogata, S., 2016. Mechanism transition and strong temperature dependence of dislocation nucleation from grain boundaries: An accelerated molecular dynamics study. Physical Review B 94, 104110.

Gangloff, R.P., Somerday, B.P., 2012. Gaseous Hydrogen Embrittlement of Materials in Energy Technologies. Woodhead Publishing Limited, Cambridge, UK.

Geng, W.T., Freeman, A.J., Wu, R., Geller, C.B., Raynolds, J.E., 1999. Embrittling and strengthening effects of hydrogen, boron, and phosphorus on a Sigma 5 nickel grain boundary. Physical Review B 60, 7149-7155.

Gerberich, W., 2012. Modeling hydrogen induced damage mechanisms in metals, in: Gangloff, R.P., Somerday, B.P. (Eds.), Gaseous Hydrogen Embrittlement of Materials in Energy Technologies. Woodhead Publishing Limited, Cambridge, UK.

Hayward, E., Fu, C.-C., 2013. Interplay between hydrogen and vacancies inα-Fe. Physical Review B 87.

Hoover, W.G., 1985. Canonical dynamics: Equilibrium phase-space distributions. Physical review. A 31, 1695-1697.

Huang, S., Chen, D., Song, J., McDowell, D.L., Zhu, T., 2017. Hydrogen embrittlement of grain boundaries in nickel: an atomistic study. NPJ Computational Materials 3, 1.

Itsumi, Y., Ellis, D.E., 1996. Electronic bonding characteristics of hydrogen in bcc iron: Part I. Interstitials. Journal of Materials Research 11, 2206-2213.

Kelchner, C.L., Plimpton, S.J., Hamilton, J.C., 1998. Dislocation nucleation and defect structure during surface indentation. Physical Review B 58, 11085-11088.

Kirchheim, R., 2010. Revisiting hydrogen embrittlement models and hydrogen-induced homogeneous nucleation of dislocations. Scripta Materialia 62, 67-70.

Kirchheim, R., Somerday, B., Sofronis, P., 2015. Chemomechanical effects on the separation of interfaces occurring during fracture with emphasis on the hydrogen-iron and hydrogen-nickel system. Acta Materialia 99, 87-98.

Kresse, G., Furthmüller, J., 1996. Efficient iterative schemes for ab initio total-energy calculations using a plane-wave basis set. Physical review B 54, 11169.

Kresse, G., Joubert, D., 1999. From ultrasoft pseudopotentials to the projector augmented-wave method. Physical Review B 59, 1758.

Li, S., Li, Y., Lo, Y.-C., Neeraj, T., Srinivasan, R., Ding, X., Sun, J., Qi, L., Gumbsch, P., Li, J., 2015. The interaction of dislocations and hydrogen-vacancy complexes and its importance for deformation-induced proto nano-voids formation in α-Fe. International Journal of Plasticity 74, 175-191.

Li, Y., Li, W., Hu, J.C., Song, H.M., Jin, X.J., 2017. Compatible strain evolution in two phases due to epsilon martensite transformation in duplex TRIP-assisted stainless steels with high hydrogen embrittlement resistance. International Journal of Plasticity 88, 53-69.

Liu, X.Y., Kameda, J., Anderegg, J.W., Takaki, S., Abiko, K., McMahon, C.J., 2008. Hydrogen-induced cracking in a very-high-purity high-strength steel. Materials Science and Engineering: A 492, 218-220.

Lynch, S.P., 2011. Interpreting hydrogen-induced fracture surfaces in terms of deformation processes: A new approach. Scripta Materialia 65, 851-854.

Möller, J.J., Bitzek, E., 2014. Comparative study of embedded atom potentials for atomistic simulations of fracture in α-iron. Modelling and Simulation in Materials Science and Engineering 22, 045002.

Macadre, A., Nakada, N., Tsuchiyama, T., Takaki, S., 2015. Critical grain size to limit the hydrogen-induced ductility drop in a metastable austenitic steel. International Journal of Hydrogen Energy 40, 10697-10703.

Martin, M.L., Fenske, J.A., Liu, G.S., Sofronis, P., Robertson, I.M., 2011a. On the formation and nature of quasi-cleavage fracture surfaces in hydrogen embrittled steels. Acta Materialia 59, 1601-1606.

Martin, M.L., Robertson, I.M., Sofronis, P., 2011b. Interpreting hydrogen-induced fracture surfaces in terms of deformation processes: A new approach. Acta Materialia 59, 3680-3687.

Martin, M.L., Somerday, B.P., Ritchie, R.O., Sofronis, P., Robertson, I.M., 2012. Hydrogen-induced intergranular failure in nickel revisited. Acta Materialia 60, 2739-2745.

Mendelev, M.I., Han, S., Srolovitz, D.J., Ackland, G.J., Sun, D.Y., Asta, M., 2003. Development of new interatomic potentials appropriate for crystalline and liquid iron. Philosophical Magazine 83, 3977-3994.

Mohtadi-Bonab, M.A., Szpunar, J.A., Razavi-Tousi, S.S., 2013. A comparative study of hydrogen induced cracking behavior in API 5L X60 and X70 pipeline steels. Engineering Failure Analysis 33, 163-175.

Nagao, A., Smith, C.D., Dadfarnia, M., Sofronis, P., Robertson, I.M., 2012. The role of hydrogen in hydrogen embrittlement fracture of lath martensitic steel. Acta Materialia 60, 5182-5189.

Nagumo, M., 2016. Fundamentals of Hydrogen Embrittlement. Springer, Singapore.

Neeraj, T., Srinivasan, R., Li, J., 2012. Hydrogen embrittlement of ferritic steels: Observations on deformation microstructure, nanoscale dimples and failure by nanovoiding. Acta Materialia 60, 5160-5171.

Nibur, K.A., Somerday, B.P., Balch, D.K., San Marchi, C., 2009. The role of localized deformation in hydrogen-assisted crack propagation in 21Cr–6Ni–9Mn stainless steel. Acta Materialia 57, 3795-3809.

Nosé, S., 1984. A molecular dynamics method for simulations in the canonical ensemble. Molecular Physics 52, 255-268.

Novak, P., Yuan, R., Somerday, B.P., Sofronis, P., Ritchie, R.O., 2010. A statistical, physical-based, micro-mechanical model of hydrogen-induced intergranular fracture in steel. Journal of the Mechanics and Physics of Solids 58, 206-226.

Oriani, R., 1987. Whitney Award Lecture-1987: hydrogen-the versatile embrittler. Corrosion 43, 390-397.

Parrinello, M., Rahman, A., 1981. Polymorphic Transitions in Single-Crystals - a New Molecular-Dynamics Method. Journal of Applied Physics 52, 7182-7190.

Perdew, J.P., Burke, K., Ernzerhof, M., 1996. Generalized gradient approximation made simple. Physical review letters 77, 3865.

Plimpton, S., 1995. Fast Parallel Algorithms for Short-Range Molecular-Dynamics. J Comput Phys 117, 1-

19.

Ramasubramaniam, A., Itakura, M., Carter, E.A., 2009. Interatomic potentials for hydrogen in α–iron based on density functional theory. Physical Review B 79.

Robertson, I.M., Martin, M.L., Fenske, J.A., 2012. Influence of hydrogen on the behavior of dislocations, in: Gangloff, R.P., Somerday, B.P. (Eds.), Gaseous Hydrogen Embrittlement of Materials in Energy Technologies. Woodhead Publishing Limited, Cambridge, UK.

Robertson, I.M., Sofronis, P., Nagao, A., Martin, M.L., Wang, S., Gross, D.W., Nygren, K.E., 2015. Hydrogen Embrittlement Understood. Metallurgical and Materials Transactions B 46, 1085-1103.

Shibata, A., Momotani, Y., Murata, T., Matsuoka, T., Tsuboi, M., Tsuji, N., 2017. Microstructural and crystallographic features of hydrogen-related fracture in lath martensitic steels. Mater Sci Tech-Lond 33, 1524-1532.

Shibata, A., Murata, T., Takahashi, H., Matsuoka, T., Tsuji, N., 2015. Characterization of Hydrogen-Related Fracture Behavior in As-Quenched Low-Carbon Martensitic Steel and Tempered Medium-Carbon Martensitic Steel. Metallurgical and Materials Transactions A 46, 5685-5696.

Shibata, A., Takahashi, H., Tsuji, N., 2012. Microstructural and Crystallographic Features of Hydrogen-related Crack Propagation in Low Carbon Martensitic Steel. Isij International 52, 208-212.

Solanki, K.N., Tschopp, M.A., Bhatia, M.A., Rhodes, N.R., 2012. Atomistic Investigation of the Role of Grain Boundary Structure on Hydrogen Segregation and Embrittlement in α-Fe. Metallurgical and Materials Transactions A 44, 1365-1375.

Song, J., Curtin, W.A., 2013. Atomic mechanism and prediction of hydrogen embrittlement in iron. Nature materials 12, 145-151.

Song, J., Curtin, W.A., 2014. Mechanisms of hydrogen-enhanced localized plasticity: An atomistic study using α-Fe as a model system. Acta Materialia 68, 61-69.

Tahir, A., Janisch, R., Hartmaier, A., 2014. Hydrogen embrittlement of a carbon segregated Σ5 (310)[001] symmetrical tilt grain boundary in α-Fe. Materials Science and Engineering: A 612, 462-467.

Takasawa, K., Ikeda, R., Ishikawa, N., Ishigaki, R., 2012. Effects of grain size and dislocation density on the susceptibility to high-pressure hydrogen environment embrittlement of high-strength low-alloy steels. International Journal of Hydrogen Energy 37, 2669-2675.

Tarzimoghadam, Z., Ponge, D., Klöwer, J., Raabe, D., 2017. Hydrogen-assisted failure in Ni-based superalloy 718 studied under in situ hydrogen charging: The role of localized deformation in crack propagation. Acta Materialia 128, 365-374.

Tiegel, M.C., Martin, M.L., Lehmberg, A.K., Deutges, M., Borchers, C., Kirchheim, R., 2016. Crack and blister initiation and growth in purified iron due to hydrogen loading. Acta Materialia 115, 24-34.

Troiano, A.R., 1960. The role of hydrogen and other interstitials in the mechanical behavior of metals. Trans ASM 52, 54-80.

Tucker, G.J., McDowell, D.L., 2011. Non-equilibrium grain boundary structure and inelastic deformation using atomistic simulations. International Journal of Plasticity 27, 841-857.

Wang, S., Martin, M.L., Robertson, I.M., Sofronis, P., 2016. Effect of hydrogen environment on the separation of Fe grain boundaries. Acta Materialia 107, 279-288.

Wang, S., Martin, M.L., Sofronis, P., Ohnuki, S., Hashimoto, N., Robertson, I.M., 2014. Hydrogen-induced intergranular failure of iron. Acta Materialia 69, 275-282.

Xie, D., Li, S., Li, M., Wang, Z., Gumbsch, P., Sun, J., Ma, E., Li, J., Shan, Z., 2016. Hydrogenated vacancies lock dislocations in aluminium. Nature communications 7, 13341.

Zhang, Z., Obasi, G., Morana, R., Preuss, M., 2016. Hydrogen assisted crack initiation and propagation

in a nickel-based superalloy. Acta Materialia 113, 272-283.

Zhu, Y., Li, Z., Huang, M., Fan, H., 2017. Study on interactions of an edge dislocation with vacancy-H complex by atomistic modelling. International Journal of Plasticity 92, 31-44.